\documentstyle[prb,aps]{revtex}

\begin{document}


\newcommand{\BEQ}{\begin{equation}}    
\newcommand{\BEA}{\begin{eqnarray}}
\newcommand{\EEQ}{\end{equation}}      
\newcommand{\EEA}{\end{eqnarray}}
\newcommand{\eps}{\epsilon}                      
\newcommand{\lmb}{\lambda}                       
\newcommand{\sig}{\sigma}                        
\newcommand{\vph}{\varphi}                       
\newcommand{\ups}{\Upsilon}                      
\newcommand{\rar}{\rightarrow}                   

\input epsf.sty
\twocolumn[\hsize\textwidth\columnwidth\hsize\csname %
@twocolumnfalse\endcsname
 
\draft

\widetext


\title{Finite-size scaling in thin Fe/Ir(100) layers}

\author{Malte Henkel,$^a$ St\'ephane Andrieu,$^a$ Philippe Bauer$^b$ and 
Michel Piecuch$^a$}
\address{$^a$Laboratoire de Physique des Mat\'eriaux\cite{URA},
Universit\'e Henri Poincar\'e Nancy I, B.P. 239, \\
F - 54506 Vand{\oe}uvre l\`es Nancy Cedex, France}
\address{$^b$LMIT, 4 Place Thanadin, F- 25200 Montb\'eliard, France}

\date{14 November 1997/ 23 February 1998}
\maketitle
  
\begin{abstract}

The critical temperature of thin Fe layers on Ir(100) is measured through
M\"o{\ss}bauer spectroscopy as a function of the layer
thickness. From a phenomenological finite-size scaling analysis, 
we find an effective shift exponent $\lambda = 3.15\pm 0.15$, 
which is twice as large as the value
expected from the conventional finite-size scaling prediction $\lambda=1/\nu$,
where $\nu$ is the correlation length critical exponent.  
Taking corrections to finite-size scaling into account,
we derive the effective shift exponent 
$\lambda=(1+2\Delta_1)/\nu$, where $\Delta_1$ describes the 
leading corrections to scaling. For the $3D$ Heisenberg
universality class, this leads to $\lmb=3.0\pm 0.1$, in agreement with the
experimental data. 
Earlier data by Ambrose and Chien on the effective shift exponent in
CoO films are also explained. 
\end{abstract}


\phantom{.}
]

\narrowtext

The study of finite-size effects in critical phenomena has for a long time
been an issue for theoretical studies, 
see Refs. \onlinecite{Barb83,Priv90,Chri93}
for reviews. Only recently, due to important advances in epitaxial
techniques necessary for the precise preparation of thin films, has this
area become accessible to experiments.\cite{Prin81} 
Presently, experiments mainly study the
finite-size shift of the critical temperature $T_c(n)$ of a thin film of
$n$ layers, as phenomenologically described by the shift exponent $\lambda$. 

Practically, two ways of defining the shift exponent have been used.
Traditionally \cite{Barb83}, one measures the shift 
of $T_c(n)$ with respect to the
bulk critical temperature $T_{c,b}=T_c(\infty)$ and sets
\BEQ \label{eq:deltaT}
\delta T := \left( T_c(\infty) - T_c(n) \right)/T_c(\infty) \sim n^{-\lmb}
\EEQ
in the limit when $n\rar\infty$. This defines the exponent $\lmb$. 
Standard finite-size scaling theory relates this to the correlation 
length exponent $\nu$, viz. $\lmb=1/\nu$ 
(Refs. \onlinecite{Barb83,Priv90,Chri93}).
Alternatively, as advocated for example in Ref. \onlinecite{Huan93}, 
one may also define
\BEQ \label{eq:DeltaT}
\Delta T := \left( T_c(\infty) - T_c(n) \right)/T_c(n) \sim n^{-\lmb'}
\EEQ
which defines the exponent $\lmb'$. Again, the limit $n\rar\infty$
is implied and from standard finite-size scaling\cite{Barb83,Priv90,Chri93}
it follows that $\lmb'=1/\nu$. Empirically, using
$\Delta T$ rather than $\delta T$ appears advantageous, since measured
values for $T_c(n)$ coincide with a power 
law for a larger range of values
of $n$ for $\Delta T$ than for $\delta T$. 

Shift exponents have been extracted successfully from either
(\ref{eq:deltaT}) or (\ref{eq:DeltaT}) for a variety of 
systems.\cite{Huan93,Lede93,Ambr96,Meht97,Ball90,Elme94,Farl93} 
For a magnetic material adsorbed as a thin film on some non-magnetic substrate,
the interaction of the magnetic moments with the substrate may restrict
the degrees of freedom of the microscopic magnetic moments. 
Examples of thin films of magnetic materials or liquid helium 
realizing the $3D$ universality
classes of the Ising model\cite{Lede93,Ambr96}, 
the XY model\cite{Meht97} and the
Heisenberg model\cite{Huan93,Ball90}, as well as the $2D$ Ising 
model,\cite{Elme94} have been constructed and analysed. The values
of the shift exponent agree with those expected from $\lmb=\lmb'=1/\nu$ 
to within a few per cent in all these cases, well within the experimental
error bars. 
In practice, this is an important alternative to measure true critical
exponents, since the critical region in finite-size scaling studies
is much larger than in thermal measurements of the bulk 
quantities, as is well known.\cite{Barb83,Priv90,Chri93}

However, the use of
(\ref{eq:DeltaT}) has been criticized in Ref. \onlinecite{Ambr96}, on the 
grounds that measurements on the N\'eel temperatures in CoO/SiO$_2$ multilayers
yielded\cite{Ambr96} $\lmb=1.6\pm 0.1$, but $\lmb'=3.4\pm 0.3$, 
which are clearly different. It was concluded in Ref. \onlinecite{Ambr96} 
that use of  eq.~(\ref{eq:DeltaT}) were best avoided since the results for 
$\lmb'$ had no interpretation. 

In this letter, we shall re-examine 
this point and shall show how corrections to finite-size scaling may
be invoked to provide a simple interpretation of those cases where a large 
value of $\lmb'$ is measured.  
At the same time, we shall 
report data on the critical temperature $T_c(n)$ as
obtained from M\"o{\ss}bauer spectroscopy of thin layers of Fe on Ir(100),
which can be analysed and understood in the same way. 
As a bonus, the correction-to-scaling
exponent $\Delta_1$ can be measured experimentally. 

We begin describing the experimental procedure. 
As M\"o{\ss}bauer spectroscopy is not sufficiently sensitive to get some
signal on so thin layers, Fe/Ir superlayers were prepared, 
where the Fe/Ir bilayer is
repeated 20 times. In this study, the Fe thickness was varied 
from 2 to 8 atomic planes
and the Ir thickness is kept constant and equal to $1.5 {\rm nm}$ from 
one superlattice
to another. The Fe/Ir(100) superlattices were prepared by 
molecular beam epitaxy
(MBE). All the details of the sample preparation are 
given in Ref. \onlinecite{And95}.
The Fe/Ir system is a good candidate to investigate the 
magnetic properties of
ultrathin films, since the Fe on Ir {\it and} the Ir on 
Fe growth mode \cite{And95a}
is an atomic layer-by-layer growth process -- also called $2D$ growth. Indeed,
abrupt and flat Fe/Ir and Ir/Fe interfaces are obtained as shown by high
resolution microscopy.\cite{Sno96} 
Moreover, the occurrence of $2D$ growth allows
us to control accurately the atomic quantity necessary to complete 
one atomic layer
by using electron diffraction. Owing to this particular behaviour, 
we really deal
with iron slabs constituted of exactly $n$ atomic planes. 
A correct analysis of the
variation of Fe magnetic properties with the number $n$ of 
deposited atomic planes
is thus possible. In this connection, M\"o{\ss}bauer 
spectrometry appears well suited.
This technique provides a measurement of the so-called 
hyperfine interactions which
occur between a resonant nucleus and its electronic
surrounding. Among these, the magnetic 
hyperfine interactions manifests itself by
a Zeeman splitting of the nuclear spin 
state when a steady local magnetic field acts on
a resonant nucleus. In case of the $^{57}$Fe 
in ferromagnetic iron, the magnetic 
hyperfine field arises mainly from the core 
polarization due to the $3d$ moment. The net
$s-$spin density 
at the nucleus is proportional,
but opposite, to the atomic on-site magnetic moment 
$\vec{M}$. So, the room 
temperature M\"o{\ss}bauer spectrum of standard bcc 
iron consists of six line patterns;
the measured energy positions provide the magnitude of the hyperfine field 
${\it Bhf} = 33$ T. Above the magnetic ordering temperature, 
due to fast thermal
atomic spin flip in the paramagnetic regime, 
${\it Bhf}=0$: the splitting vanishes and
the spectrum exhibits only one line 
(see for example Ref. \onlinecite{Blu68}). 
Usually, in case of normal bulk iron 
ferromagnets, the transition takes place within
a narrow temperature range and the thermal 
variation ${\it Bhf} = {\it Bhf}(T)$ goes 
along with $M(T)$.\cite{Pres62} 

\begin{figure}
\centerline{\epsfxsize=3.75in\epsfbox
{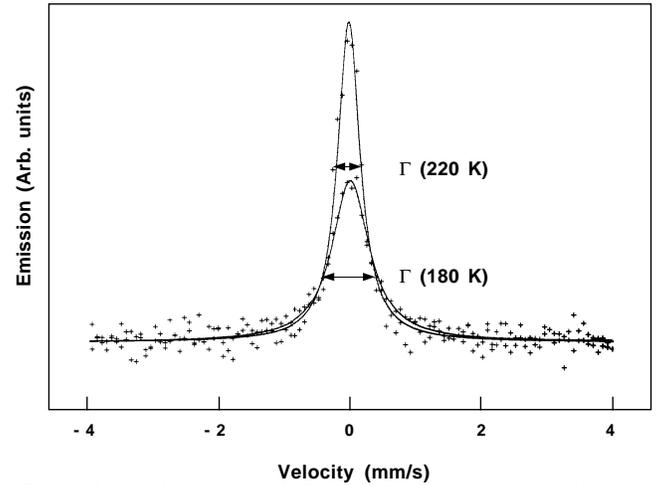}}
\caption{Example of the M\"o{\ss}bauer line broadening when the temperature
is decreased, for a $6 {\rm ML}$ $^{57}$Fe superlattice. Down to $T=T_c$, the
FWHM $\Gamma$ is constant. When $T_c$ is reached, $\Gamma$ begins to
increase.\label{fig:spektra}
}
\end{figure}

As only $2\%$ of $^{57}$Fe is present in natural Fe, 
the MBE chamber was equipped with
a $92\%$ enriched $^{57}$Fe source, in order to get a 
sufficient amount of $^{57}$Fe
in the Fe layers of the superlattices. 
M\"o{\ss}bauer spectrometry was performed in
backscattered mode by detecting conversion 
electron after resonant absorption of
$14.4$ keV $\gamma$ rays emitted by a $^{57}$Co 
source. Source drive and data storage 
were conventional, the low-temperature electron 
detector was a circular microchannel
plate housed in a home-made cryostat \cite{Bau96}. 
For all investigated Fe/Ir(100)
superlattices, the weakening and vanishing of 
the Zeeman splitting was observed to
take place in a rather wide temperature range, 
compared to the collapse which occurs
for standard bcc iron when heating up. Thus, 
ordering temperatures of the superlattices
are here defined as the temperature for which 
the {\em onset} of a line broadening takes place in
the paramagnetic spectrum when cooling down 
from room temperature. Such a determination is illustrated in
figure~\ref{fig:spektra} with typical M\"o{\ss}bauer spectra
recorded above and below the critical point. We have checked by
neutron diffraction that, at least for the Fe thickness range where the
Fe is still strained by Ir, there is no coupling of two Fe layers mediated 
the intervening Ir layer. The measured $T_c(n)$ are thus the critical
temperatures of a single Fe layer. For each value of $n$, a single
specimen was made. The quoted errors in $T_c(n)$ correspond to the
steps in temperature used in the scanning of the M\"o{\ss}bauer spectra. 
For $n=2,\ldots,6$ deposited iron monolayers, the ordering 
temperatures are given in table~\ref{exptab}.
For 8 monolayers, the spectrum exhibits splitting at room temperature.

\begin{figure}
\centerline{\epsfxsize=3.25in\epsfbox
{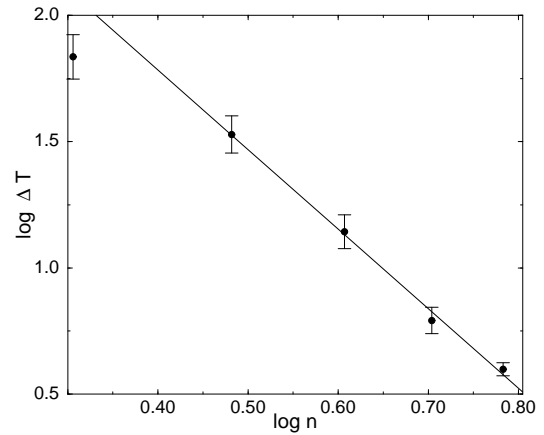}}
\caption{Shift $\Delta T$ from eq.~(\ref{eq:DeltaT}) 
of the critical temperature of the Fe/Ir(100) system as a function of
the number $n$ of Fe monolayers, for $n=2,\ldots,6$. The curve is the
power-law fit $\Delta T(n) = 1069\, n^{-3.15}$.
\label{fig:Gerade}
}
\end{figure}

The dependence of the shift $\Delta T(n)$ on $n$, calculated using the value
$T_{c,b}=1043 \mbox{\rm K}$ for the critical point of bulk iron, is shown in 
figure~\ref{fig:Gerade}. From at least $n=3$ monolayers on, 
the data are very well described in terms of a power law.\cite{kleinN}
Independently of any further theoretical interpretation, our data show that the
system does lie inside the finite-size scaling region. Therefore, the analysis
of the systematic variation of the critical point $T_c(n)$ with the number of 
monolayers $n$ in terms of a phenomenological shift exponent is 
justified. We find
\BEQ \label{eq:Lambda}
\lmb_{\rm eff}' = 3.15 \pm 0.15 
\EEQ
This is about twice the value expected from $\lmb'=1/\nu$ 
in the $3D$ Heisenberg model, see table~\ref{tab1}. 
Since in the present setting, we explore
the transition from a $3D$ bulk system to a $2D$ film, we expect that this
cross-over due to finite-size effects should be described in terms of
the exponents of the $3D$ universality classes. 

We now come back to the problem of explaining the value of $\lmb'$
found. Our starting point is the theory of finite-size 
scaling.\cite{Barb83,Priv90,Chri93} Given that the
values of $T_c(n)$ (for the values of $n$ accessible to experiment)
are much lower than the bulk $T_{c,b}$, it appears sensible to
consider not only the leading effects of finite-size scaling but to
take finite-size corrections into account as well. For the magnetization,
measured on a film of a thickness of $n$ monolayers, we 
expect \cite{Barb83,Priv90,Chri93}
\BEQ \label{eq:FSSM}
M_n \simeq n^{-\beta/\nu} Z\left( t n^{1/\nu}, u n^{y_3}\right)
\EEQ
where $t\sim T_c(n)-T_{c,b}$, $\beta$ and $\nu$ are the conventional
magnetization and correlation length exponents, $u$ stands for an
irrelevant scaling variable which parametrizes the leading finite-size
corrections and $y_3 < 0$ is the associated exponent. Finally, $Z=Z(z_1,z_2)$
is a scaling function, which is independent of the layer thickness $n$.  

The pseudocritical point $T_c(n)$ (and thus an associated 
$t_n$ as well) is determined
from the vanishing of the magnetization, viz. $M_n(t_n)=0$. This implies
\BEQ \label{eq:tncond}
Z\left( t_n n^{1/\nu}, u n^{y_3} \right) = 0
\EEQ
From this, we have to derive the scaling of $t_n$ as a function of $n$. 
In particular, from figure~\ref{fig:Gerade} we expect that phenomenologically
\BEQ \label{eq:tnlmb}
t_n \sim \tau \, n^{-\lmb}
\EEQ
where $\lmb=\lmb_{\rm eff}$ is the effective shift exponent and $\tau$
is a constant. (For the purpose of this discussion, the distinction 
between $\lmb$ and
$\lmb'$ as defined in eqs.~(\ref{eq:deltaT},\ref{eq:DeltaT}) is unnecessary.) 
When finite-size effects are small, we can simply set $u=0$
in (\ref{eq:tncond}) and then recover the standard 
result\cite{Barb83,Priv90,Chri93} $\lmb=\lmb'=1/\nu$. Here, we want to discuss 
how large finite-size corrections might affect the value of $\lmb_{\rm eff}$.

For $n-1$ monolayers, we can find, in the same way, $t_{n-1}$ from
$Z( t_{n-1} (n-1)^{1/\nu}, u (n-1)^{y_3} ) = 0$. For $n$ large enough,
we can expand the arguments of $Z$ and get, to leading order in $1/n$
\BEA 
\lefteqn{ Z\left( t_n n^{1/\nu}, u n^{y_3} \right) -y_3 u n^{y_3-1} Z_2} 
\nonumber \\
&+& \left( -\frac{1}{\nu} t_n n^{1/\nu-1} + 
\left( t_{n-1}-t_n\right) n^{1/\nu} \right) Z_1
\simeq 0 \label{eq:fssanal}
\EEA
where $Z_{1,2}=\partial Z/\partial z_{1,2}(t_n n^{1/\nu}, u n^{y_3})$.
To leading order in $1/n$, they can be considered as constants. The
first term in (\ref{eq:fssanal}) 
vanishes due to the condition (\ref{eq:tncond}), used to
determine $t_n$. In addition, from (\ref{eq:tnlmb}) we have 
$t_{n-1}-t_n \simeq -\lmb\tau n^{-\lmb-1}$.
Inserting into (\ref{eq:fssanal}), we find 
\BEQ \label{eq:lmbtau}
\lmb = \frac{1}{\nu} - y_3 \;\; , \;\; \tau = - \frac{Z_2}{Z_1}
\frac{y_3}{1/\nu +\lmb} u
\EEQ
and, recalling that $y_3<0$, it is already clear that the effective
shift exponent may have values larger than the usually expected $1/\nu$.
The same result for $\lmb$ had been obtained previously from an analysis
of the finite-size scaling of the correlation length.\cite{Priv83} 

The preceding discussion can be generalized to show that there is a whole
series of correction terms to finite-size scaling,\cite{Priv83,first} 
each of them
with its proper value for the effective shift exponent 
$\lmb_{\rm eff} = 1/\nu - k y_3$,
where $k=0,1,2,\ldots$. It depends on the (system-dependent and non-universal)
value of $u$ whether the leading correction is enough to describe the data
or if several correction terms must be taken into account. 
For rather thick films, finite-size effects should
be unimportant and thus $k=0$, which reproduces the standard result
$\lmb=1/\nu$. For thinner films, one might find a cross-over into a
regime described by a non-zero value of $k$, where $k=1$ corresponds
to the first-order calculation given above. Higher values of $k$
apply when the lower order terms vanish.\cite{second}  

Concerning the value of $y_3$, we now relate it to existing theoretical 
predictions\cite{Zinn89} for the
three-dimensional $O(n)$ model. In these models,
corrections to scaling are described in terms of the exponent $\Delta_1$,
viz. $M(t) \sim t^{\beta} ( 1+ a t^{\Delta_1} + \ldots)$, where $a$ is a
constant. Comparing with the finite-size scaling form eq.~(\ref{eq:FSSM}),
then formally leads to $|y_3|=\Delta_1/\nu$. The scaling operator
$\psi$, which generates these leading-order corrections to scaling, is
well understood in a field theory setting of the $O(n)$ model. 
In particular, it is known \cite{Zinn89} that $\psi$ is {\em even} under
spin reversal. On the other hand, the scaling operator $\sigma$ which
corresponds to the magnetization is {\em odd} under spin reversal. 
The amplitude $\tau$ of the first-order term calculated in (\ref{eq:lmbtau})
is proportional to the expectation value $\langle \psi \sigma\rangle_{0}$, to 
be evaluated {\em at} the critical point. By symmetry, this quantity
vanishes. Thus, phenomenologically, we expect the finite-size data of the
magnetization 
to be determined by the second-order corrections, with $k=2$. We therefore
arrive at the prediction, expected to be valid in cases where finite-size
corrections are important
\BEQ \label{eq:final}
\lmb' = \lmb_{\rm eff} = \frac{1}{\nu} \left( 1 + 2 \Delta_1 \right)
\EEQ

In table~\ref{tab1}, we collect the field-theoretical predictions for the
exponents $\Delta_1$, $1/\nu$ and $(1+2\Delta_1)/\nu$. The values quoted
are the mean values of those compiled in Ref. \onlinecite{Zinn89} from
$\varepsilon$-ex\-pan\-sion and from the resummed perturbation series for the  
three-dimensional $O(n)$ model. 

We now compare (\ref{eq:final}) to the experimental data. 
First, consider
the case of Fe/Ir(100). We expect that the ferromagnetic transition 
of Fe is within the Heisenberg universality class. Our result,
given in eq.~(\ref{eq:Lambda}), is in good agreement with the expected
value $\lmb'=3.0\pm 0.1$. Second, 
we consider the CoO/SiO$_2$ system.\cite{Ambr96} CoO is known to
be an antiferromagnet with localized moments and the transition is
expected to be in the Ising universality class. 
The experimental result, obtained using eq.~(\ref{eq:DeltaT}), 
$\lmb_{\rm eff}=3.4\pm 0.3$, 
is in agreement with the expectation $\lmb' = 3.2\pm 0.1$ from 
table~\ref{tab1}. Our
prediction (\ref{eq:final}) is thus clearly confirmed for two
distinct universality classes. To our best knowledge, 
this is the first time that the correction-to-scaling exponent
$\Delta_1$ has been measured experimentally. Accepting the theoretical
values of $\nu$, we find $\Delta_1\simeq0.57(9)$ and $0.61(6)$ for the Ising 
and Heisenberg universality classes, respectively. Our findings are well
consistent with the theoretical predictions obtained from field
theory, see table~\ref{tab1}. 

Some more comments are in order. The consistency of the measured shift 
exponents with $\lmb=\lmb'=1/\nu$ in most of the systems studied so 
far\cite{Huan93,Lede93,Ambr96,Meht97,Ball90,Elme94,Farl93} indicates
that, usually, finite-size corrections are apparently not very important.
On the other hand, in the CoO system,\cite{Ambr96} the shift exponent
had been measured using {\em both}
$\delta T$ and $\Delta T$. While in the second case, we have shown that
finite-size corrections are needed for the proper interpretation of the 
value of $\lmb'$, the first case yields a value 
$\lmb=1.6\pm 0.1$, close to the expected $1/\nu$. We stress that it
depends on the non-universal value of the coupling $u$ (which cannot be
predicted from our purely phenomenological analysis), whether a clear
power law can be observed for one or other of the shift 
eqs.~(\ref{eq:deltaT},\ref{eq:DeltaT}), if any, and in what regime of effective
exponents the data will finally fall.

In conclusion, we have presented data on the critical temperature of thin
Fe/Ir(100) layers. The thickness-dependence of $T_c(n)$ has been analysed
using the phenomenological theory of finite-size scaling. Finite-size 
corrections were shown to play an important role in the interpretation
of the effective exponent value $\lmb_{\rm eff}$, leading to the prediction
(\ref{eq:final}), resolving an argument on the proper extraction of
critical exponents from experimental data. 
Our data for Fe/Ir(100) and data\cite{Ambr96} on CoO/SiO$_2$ are in agreement
with the theoretical predictions coming from the $O(n)$ model. This has
allowed to experimentally confirm the field-theoretic prediction of 
the value of the correction-to-scaling 
exponent $\Delta_1$ for the $3D$ Ising and Heisenberg universality classes.

\vspace{-2mm}


\begin{table}
\begin{tabular}{|c|ccccc|} 
$n$ & 2 & 3 & 4 & 5 & 6 \\ \hline
$T_c$ & $15 \pm 3$ & $30\pm 5$ & $70\pm 10$ & $145\pm 15$ & $210\pm 10$ \\
\end{tabular}
\caption{Transition temperatures $T_c(n)$ in Kelvin obtained from the
onset of line
broadening in the paramagnetic spectrum when cooling the system, 
as a function of the number $n$ of Fe monolayers. \label{exptab}
}
\end{table}
\begin{table}
\begin{tabular}{|lc|ccc|} 
model & $n$ & $\Delta_1$ & $1/\nu$ & $(1+2\Delta_1)/\nu$\, \\ \hline
Ising & 1 & 0.50(2) & 1.586(4) & 3.17(6) \\
XY    & 2 & 0.53(2) & 1.492(7) & 3.07(11) \\
Heisenberg & 3 & 0.555(25) & 1.413(9) & 2.98(13) \\ 
\end{tabular}
\caption{Field theory predictions{\protect \cite{Zinn89}} 
from the three-di\-men\-sio\-nal
$O(n)$ vector model of some critical exponents. The numbers in brackets
estimate the error on the last given digits. \label{tab1}
}
\end{table} 

\begin{references}
\bibitem{kleinN} For the film with $n=2$, one should expect sizable systematic 
corrections to the leading finite-size scaling law (\ref{eq:DeltaT}). This 
is well known from numerous computer simulations.\cite{Barb83,Priv90}
\bibitem{first} These terms come from the non-linear scaling fields which
enter into the scaling functions, e.g. $g_u \sim u + {\rm cste}\, u^2$. 
They can
be obtained by formally developing $Z(z_1,z_2)$
in $z_2=u n^{y_3}$ around the critical point $z_1^*, z_2^*=0$, 
with coefficients proportional to the 
expectation values $\langle \sig\psi^k \rangle_0$.
\bibitem{second} For example, if in (\ref{eq:fssanal}) $Z_2=0$, one has to
go to the second order terms in $z_2$. Phenomenologically, this leads to the
same result for $\lmb$ as before, with $y_3$ replaced by $2 y_3$.
\end{references}
\end{document}